\documentclass[12pt]{article}
\pdfoutput=1
\usepackage[sumlimits,intlimits,namelimits]{amsmath}
\usepackage{amssymb}
\usepackage[english]{babel}
\usepackage{bbm}
\usepackage{parskip}
\usepackage[para]{footmisc}
\usepackage[margin=1cm]{caption}

\usepackage{graphicx}          

\usepackage{cite}

\usepackage[makeroom]{cancel}
\usepackage[usenames,dvipsnames]{color}
\usepackage[dvipsnames]{xcolor}

\usepackage[bordercolor=black,backgroundcolor=yellow,linecolor=black,
textsize=scriptsize,shadow]{todonotes}

\newcommand{\Bala}{\cite{Balasubramanian:2013ux}}

\newcommand{\AdS}{$\text{AdS}_{5}$}

\newcommand{\sn}{\mathfrak{sn}}
\newcommand{\cn}{\mathfrak{cn}}
\newcommand{\dn}{\mathfrak{dn}}
\newcommand{\am}{\mathfrak{am}}

\renewcommand{\Re}{\text{Re}}
\renewcommand{\Im}{\text{Im}}

\newcommand{\mS}{\mathcal{S}}

\newcommand{\mK}{\mathcal{K}}

\newcommand{\mO}{\mathcal{O}}

\newcommand{\mP}{\mathcal{P}}

\newcommand{\Poincare}{Poincar\'e}

\global\parskip 6pt

\newcommand{\qed}{\nobreak \ifvmode \relax \else\ifdim\lastskip<1.5em 
\hskip-\lastskip\hskip1.5em plus0em minus0.5em \fi \nobreak\vrule height0.75em 
width0.5em depth0.25em\fi}

\newcommand{\be}{\begin{eqnarray}}
\newcommand{\ee}{\end{eqnarray}}

\def\>{\rangle}
\def\<{\langle}

\newcommand{\executeiffilenewer}[3]{%
\ifnum\pdfstrcmp{\pdffilemoddate{#1}}%
{\pdffilemoddate{#2}}>0%
{\immediate\write18{#3}}\fi%
}

\graphicspath{{pics/}}

\usepackage{hyperref}

\begin{document}

\begin{titlepage}
\vspace*{.5cm}
\begin{center}
{\Large{\bf Discrete scale invariance in holography revisited}} \\[.5ex]
\vspace{1cm}
Mario Flory\footnote{mflory@th.if.uj.edu.pl},
\vspace{0.5cm}\\
\em
Institute of Physics, Jagiellonian University, \\
\L{}ojasiewicza 11, 30-348 Krak\'ow, Poland
\end{center}
\vspace*{1cm}
\begin{abstract}
In 2013, Balasubramanian presented a $5+1$ dimensional holographic toy model that allows for an exact solution to Einstein's equations in the bulk in which the isometries of \AdS\ appear to be broken to an isometry group describing a discretely scale invariant and \Poincare\ invariant setup \cite{Balasubramanian:2013ux}. In this paper, we investigate this solution in more detail. By analytically solving the Killing equations, we prove that the full \AdS\ isometry group is still present, although in a somewhat hidden way. We will also comment on the prospects of finding other holographic bottom up toy models which allow for solutions with discrete scale invariance or scale invariance without conformal invariance in the future. 
\end{abstract}
\vspace*{.25cm}
\end{titlepage}

\tableofcontents
\thispagestyle{empty}

\section{Introduction: Discrete scale invariance}
\label{sec::Intro}

\textit{Scale invariance} is the property of a mathematical or physical system in which observables are well-behaved under scale transformations $x\rightarrow \lambda x$. For example, for a scale invariant observable $\mO_{SI}(x)$, we demand the law
\begin{align}
\mO_{SI}(x)=\mu(\lambda)\mO_{SI}(\lambda x)
\label{si}
\end{align}
to hold for \textit{any} $\lambda\in\mathbb{R}^+$ and some $\mu(\lambda)=\lambda^\alpha$ with \textit{critical exponent} $\alpha\in\mathbb{R}$. We say that a system exhibits \textit{discrete scale invariance} (DSI) if the equation
\begin{align}
\mO_{DSI}(x)=\mu(\lambda_0)\mO_{DSI}(\lambda_0 x)
\label{dsi}
\end{align}
\textit{only} holds for a specific scale $\lambda_0\in\mathbb{R}^+$ and the related scales $\lambda_0^m,\ m\in\mathbb{Z}$. Equation \eqref{dsi} can be solved by
\begin{align}
\mO_{DSI}(x)\propto x^\alpha,\ \ \alpha=-\frac{\log \mu}{\log \lambda_0}+i\frac{2\pi n}{\log \lambda_0},\ \ n\in\mathbb{Z}.
\label{DSIsol}
\end{align}
The hallmark signatures of DSI are hence \textit{complex critical exponents} such as $\alpha$ in \eqref{DSIsol} for $n\neq0$ and, because of
\begin{align}
\mO_{DSI}(x)&\propto x^\alpha
=x^{\Re(\alpha)}\left(\cos\left[\Im(\alpha)\log(x)\right]+i\sin\left[\Im(\alpha)\log(x)\right]\right),
\nonumber
\end{align}   
the appearance of \textit{log-periodic oscillations} of observables. This type of symmetry, respectively the associated signatures, can be exhibited by a variety of systems such as fractals (for example the famous Koch curve or the triadic Cantor set) \cite{Sornette:1997pb}, stock markets \cite{SORNETTE1997411,refId0,Sornette:1997pb}, earthquakes \cite{refId1,0295-5075-41-1-043,saleur:hal-00331040,Sornette:1997pb}, black hole formation 
\cite{Choptuik:1992jv,Abrahams:1993wa,Hirschmann:1994du,Hirschmann:1995qx}, perturbations of extremal black holes (see e.g.~\cite{Gralla:2017lto}), black hole/black string phase transitions \cite{Kol:2002xz,Kalisch:2017bin}, the Efimov effect \cite{Hammer:2011kg}, QFT toy models \cite{Georgi:2016tjs}, quantum gravity \cite{Calcagni:2017via}, condensed matter models \cite{0305-4470-29-13-017,PhysRevB.90.184202} and even holographic AdS/CMT models \cite{Liu:2009dm,Faulkner:2009wj,Hartnoll:2015rza,Erdmenger:2016msd,Brattan:2017yzx}. See \cite{Sornette:1997pb} for a general review. In the context of \textit{renormalisation group (RG) flows}, DSI means that instead of a fixed point, the system under investigation exhibits \textit{cyclic} RG behaviour \cite{Wilson:1970ag,Glazek:2002hq,LeClair:2002ux,LeClair:2003hj,Curtright:2011qg,Bulycheva:2014twa}.\footnote{The converse, however, is not necessarily true as cyclic RG flows can be equivalent to conformal fixed points \cite{Luty:2012ww,Fortin:2012hn}, see also the discussion in \Bala. Also, discrete scale \textit{covariance} can lead to an RG behaviour in which certain couplings are cyclic, but in which there is still a clear direction to the flow, in which for example degrees of freedom decrease monotonically \cite{Franco:2004jz,Shaghoulian:2016umj}.}

In the last years, \textit{holography} or the \textit{AdS/CFT correspondence} has proven to be an extremely useful tool to gain insight into many topics of quantum field theory (see e.g.~\cite{Ammon:2015wua}), and starting with \cite{Freedman:1999gp}, there have been a number of constructions of interesting types of holographic RG flows, some of them exhibiting chaotic \cite{Franco:2004jz,Shaghoulian:2016umj}, "Boomerang" \cite{Donos:2017ljs,Donos:2017sba}, or otherwise exotic behaviour \cite{Kiritsis:2016kog}.

In order to learn more about cyclic RG flows and discrete scale invariance, it might be useful to construct holographic models which exhibit this type of symmetry. As already mentioned above, there are a few holographic models \cite{Liu:2009dm,Faulkner:2009wj,Hartnoll:2015rza,Erdmenger:2016msd,Brattan:2017yzx} in which log-periodic oscillations of certain variables are known to occur. In the models \cite{Liu:2009dm,Faulkner:2009wj,Hartnoll:2015rza}, however, these log-periodic oscillations did not occur on a spacetime axis, but instead as a function of a frequency or temperature. In \cite{Erdmenger:2016msd}, log-periodic oscillations of certain observables where observed as a function of time, but only after a local quench when the system was momentarily out of equilibrium. The model of \cite{Brattan:2017yzx} exhibited Lifshitz-scaling, i.e.~anisotropic scaling between space and time. 

So in order to study \Poincare- \textit{and} discretely scale invariant systems holographically, it may be better to build a corresponding toy model from scratch by breaking the scale invariance encoded in the isometry group of AdS down to discrete scale invariance:
\begin{align}
&ds^2_{\text{AdS}}
=e^{2w/L}\left(-dt^2+d\vec{x}^2\right)+dw^2
\label{AdS}
\\
\rightarrow
&ds^2
=e^{2w/L+f(w)}\left(-dt^2+d\vec{x}^2\right)+dw^2
\label{periodicAdS}
\end{align} 
where $f(w)\neq const.$ is a periodic function in $w$. However, as pointed out in \cite{Balasubramanian:2013ux}, the holographic $c$-theorem \cite{Freedman:1999gp} provides a simple argument that the symmetry breaking \eqref{periodicAdS} can only be possible when the \textit{null energy condition (NEC)}\footnote{This condition posits that $T_{\mu\nu}k^\mu k^\nu\geq0$ where $T_{\mu\nu}$ is the energy-momentum tensor and $k^\nu$ is any null-vector field. Assuming the validity of Einstein's equations, this implies $R_{\mu\nu}k^\mu k^\nu\geq0$ where $R_{\mu\nu}$ is the Ricci tensor. See \cite{Curiel:2014zba} for a review on energy conditions.} is violated. The reasoning behind this argument is that the $c$-theorem of \cite{Freedman:1999gp} is based on the result that in a domain-wall ansatz of the form   
\begin{align}
ds^2_{\text{DW}}
=e^{2A(w)}\left(-dt^2+d\vec{x}^2\right)+dw^2,
\label{domainwall}
\end{align}
the NEC implies $A''(w)\leq0$, ruling out any periodic behaviour in \eqref{periodicAdS} \cite{Balasubramanian:2013ux}. As suggested in \cite{Balasubramanian:2013ux}, it might be possible, however, to circumvent this apparent no-go theorem by adding one or multiple warped extra dimensions to the domain-wall ansatz of \eqref{domainwall}. For the simplest case of one extra compact dimension, the new ansatz then takes the form
\begin{align}
ds^2_{\text{DSI}}=&\left(g_{\text{DSI}}\right)_{\mu\nu}dx^\mu dx^\nu
\label{genAnsatz}
\\
=&e^{2C(w,\theta)}\left(e^{2w/L}\left(-dt^2+d\vec{x}^2\right)+dw^2\right)
+e^{2B(w,\theta)}\left(d\theta+A(w,\theta)dw\right)^2
\nonumber
\end{align}
with the functions $A(w,\theta)$, $B(w,\theta)$ and $C(w,\theta)$ being periodic in $w$ and $\theta$. In \cite{Balasubramanian:2013ux}, both a top-down and a bottom-up model where presented in which an exact solution to Einstein's equations takes a form similar to \eqref{genAnsatz} while satisfying the NEC.\footnote{See \cite{Nakayama:2011zw,Nakayama:2013is} for other holographic models of discrete scale invariance in which, however, "a strong form of the NEC" is violated.} One may wonder whether a physical dual field theory, described holographically by such a gravitational system and, as a consequence of DSI, exhibiting a cyclic RG flow behaviour, can exist in the light of field theory results such as the $c$-theorem for two dimensional field theories \cite{Zamolodchikov:1986gt} or the $a$-theorem for four dimensional field theories \cite{Komargodski:2011vj} (see also \cite{Luty:2012ww}). In \Bala, it was argued that the existence of a four dimensional dual theory would not violate the $a$-theorem if this theory does not flow from a UV fixed point perturbed by a marginal or relevant operator, i.e.~if its ostensible UV-completion is not a four dimensional CFT, but for example a higher dimensional, non-Lorentz invariant, or lattice theory. See \Bala\ for further discussion of this point. On the other hand, if it \textit{where} possible to show on the field theory side that such theories cannot exist under any circumstance, the existence of a holographic model would, by the logic of \cite{Nakayama:2009qu,Nakayama:2009fe}, indicate that this bulk model falls into the "forbidden landscape" of holography. Hence, this topic warrants further investigation in any case.

In section \ref{sec::Sol}, we will summarise the bottom-up model of \cite{Balasubramanian:2013ux} and its solution exhibiting discrete scale invariance. In section \ref{sec::Killing} we will analytically solve the Killing equation on this background, and show that there are 15 linearly independent Killing vector fields which form a full conformal algebra. This proves that the bottom-up solution found in \cite{Balasubramanian:2013ux} retains the full isometry group of \AdS\ in a hidden way, as was already suspected in \cite{Nakayama:2013is}. The possibility of finding bulk metrics that implement scale invariance without conformal invariance will be investigated in section \ref{sec::scale}. We will discuss our results and give conclusions in section \ref{sec::conc}.

\section{A holographic bottom up model for discrete scale invariance}
\label{sec::Sol}

The bottom-up model studied in \Bala\ is based on the idea of coupling Einstein-Hilbert gravity in $5+1$ dimensions to an axion-like scalar field $\chi$:\footnote{We include a factor $1/2$ which, in our opinion, is missing in \Bala.}
\begin{align}
\mS=\frac{1}{2\kappa}\int d^6x \sqrt{-g}\left(R-\frac{1}{2}\left(\frac{1}{2}\left(\partial\chi\right)^2+V(\chi)\right)\right).
\label{action}
\end{align}
The cosmological constant term is absorbed in the definition of the scalar potential, which reads
\begin{align}
V(\chi)=-\frac{8}{L^2}\frac{2-\alpha^2}{1-\alpha^2}+\frac{12}{L^2}\frac{\alpha^2}{1-\alpha^2}\cos\left(\frac{\chi}{\sqrt{2}}\right).
\label{Potential}
\end{align}
Clearly, the field values of $\chi$ are identified with a periodicity $\chi\sim\chi+2\pi\sqrt{2}$. The action \eqref{action} implies Einstein's equations
\begin{align}
R_{\mu\nu}-\frac{1}{2}Rg_{\mu\nu}=T_{\mu\nu}=\frac{1}{2}\left(\partial_{\mu} \chi\partial_{\nu} \chi-g_{\mu\nu}\left(\frac{1}{2}\left(\partial\chi\right)^2+V(\chi)\right)\right)
\label{EinsteinEq}
\end{align}
to hold, as well as the equation of motion of the scalar field. It is noteworthy that because of the specific form that the energy-momentum tensor $T_{\mu\nu}$ takes on the right-hand-side of \eqref{EinsteinEq}, any metric $g_{\mu\nu}$ that solves \eqref{EinsteinEq} for any $\chi$ and $V(\chi)$ automatically satisfies the NEC.  

In \Bala, it was shown that for the model \eqref{action}, the ansatz \eqref{genAnsatz} of combining a $4+1$ dimensional domain-wall spacetime and a circular extra dimension $\left(\theta\sim\theta+2\pi\right)$ in a warped way leads to a solution to the equations of motion of the form
\begin{align}
ds^2_{\text{sol}}=\left(g_{\text{sol}}\right)_{\mu\nu}dx^\mu dx^\nu=ds^2_{\text{DSI}}\big|_{C(w,\theta)=...,\ B(w,\theta)=...,\ A(w,\theta)=...}
\label{specAnsatz}
\end{align}
with the functions
\begin{align}
e^{2C(w,\theta)}=&1-\alpha^2\big(\sn\left(w/h,\alpha^2\right)\cn\left(N_a \theta,\alpha^2\right)
+\cn(w/h,\alpha^2)\sn(N_a \theta,\alpha^2)\big)^2,
\label{C}
\\
e^{2B(w,\theta)}=&N_a^2 L^2 \left(1-\alpha^2\right)\dn\left(N_a \theta,\alpha^2\right)^2 e^{-2C(w,\theta)},
\label{B}
\\
A(w,\theta)=&\frac{1}{N_a h}\frac{\dn\left(w/h,\alpha^2\right)}{\dn\left(N_a \theta,\alpha^2\right)},
\label{A}
\end{align}
and the scalar field 
\begin{align}
\chi(w,\theta)=2\sqrt{2}\left(\am\left(N_a \theta,\alpha^2\right)+\am\left(w/h,\alpha^2\right)\right).
\label{chi}
\end{align}
In equations \eqref{C}-\eqref{chi} the function $\am$ is the \textit{Jacobi amplitude}, while $\sn, \cn, \dn$ in \eqref{C}-\eqref{A} are \textit{Jacobi elliptic functions}\footnote{Here we use the same conventions as in \textit{Mathematica}, i.e.~$\sn\left(x,\alpha^2\right)= $\texttt{JacobiSN[}$x,\alpha^2$\texttt{]} for example. Also, note that \eqref{C} can be simplified using the identity $\sin\left(\am(x,m)+\am(y,m)\right)=\sn(x,m)\cn(y,m)+\sn(y,m)\cn(x,m)$.}. These elliptic functions depend on two variables, and are periodic in the first variable,
\begin{align}
\sn\left(x,\alpha^2\right)&=\sn\left(x+\mP(\alpha),\alpha^2\right),
\\ 
\cn\left(x,\alpha^2\right)&=\cn\left(x+\mP(\alpha),\alpha^2\right),
\\ 
\dn\left(x,\alpha^2\right)&=\dn\left(x+\mP(\alpha)/2,\alpha^2\right),
\label{periodicity}
\end{align} 
with a period\footnote{Of course, this period could be absorbed in a redefinition of the factors $N_a$ and $h$ in \eqref{C}-\eqref{chi}. This would however introduce an extra prefactor in \eqref{B} which is missing in \Bala.}  $\mP(\alpha)$ that is given by $4$ times the complete elliptic integral of the first kind,
\begin{align}
\mP(\alpha)=4\int_{0}^{\pi/2}\frac{dx}{\sqrt{1-\alpha^2\sin(x)^2}}.
\label{Period}
\end{align}
A few comments about the solution \eqref{specAnsatz} are in order: Firstly, due to the periodic identifications of both $\theta$ and $\chi$, we need to enforce the condition
\begin{align}
\chi(w,\theta+2\pi)=\chi(w,\theta)+2\pi\sqrt{2}n,\ \ n\in\mathbb{Z}.
\end{align}
This leads to a quantisation condition on $N_a$, and hence the axion flux. Secondly, in order to ensure the negativity of the Ricci scalar $R$ everywhere in the spacetime, we have to demand $0\leq\alpha<\sqrt{2/3}$. Thirdly, in the special case $\alpha=0$, $\dn(x,0)=1$. This means that the oscillating behaviour of the spacetime (visible e.g.~on the Ricci scalar) stops in this limit, and we expect the full isometry group of \AdS\ to be present. Also, the $\theta$-dependence of the metric drops out, leading to an additional Killing vector $\partial_\theta$. See \Bala\ for a further discussion. Fourthly, the metric \eqref{specAnsatz} has a vanishing Weyl tensor for \textit{any} $\alpha$, which may already hint at the fact that these solutions posses more symmetry than expected. We will proceed to investigate this in more detail in the next section.

\section{Killing fields and hidden symmetries}
\label{sec::Killing}

The isometry group of a spacetime metric is encoded in its Killing vector fields and their Lie-bracket algebra. A \textit{Killing field} $\mK^{\mu}$ is a vector field that satisfies the \textit{Killing equation}
\begin{align}
\nabla_{\mu}\mK_{\nu}+\nabla_{\nu}\mK_{\mu}=0,
\label{Killing}
\end{align}
which, on an $n$-dimensional spacetime, can at most have $n(n+1)/2$ linearly independent solutions. The Lie-bracket
\begin{align}
\left[\mK_{1},\mK_{2}\right]^{\mu}\equiv\mK_1^\nu\partial_{\nu}\mK_2^\mu-\mK_2^\nu\partial_{\nu}\mK_1^\mu
\label{bracket}
\end{align}
of two Killing fields $\mK_1$, $\mK_2$ will itself be a Killing field.

It is easy to show that metrics of the form \eqref{domainwall} and \eqref{genAnsatz} will retain the full set of Killing vectors of the Minkowski space $ds^2_{\text{Min}}
=-dt^2+d\vec{x}^2$. For a $3+1$ dimensional Minkowski space, this leads to 10 Killing fields forming a \Poincare\ algebra. $4+1$-dimensional AdS space \eqref{AdS} possesses $5$ additional Killing fields that extend the \Poincare\ algebra to a conformal algebra, which takes the form \cite{Freedman:2012zz}
\begin{align}
[M_{mn},M_{rs}]&=\eta_{mr}M_{sn}-\eta_{nr}M_{sm}-\eta_{ms}M_{rn}+\eta_{ns}M_{rm}
\label{MM}
\\
[P_m,M_{ns}]&=2\eta_{m[n}P_{s]}
\label{PM}
\\
[D,P_m]&=P_{m}
\label{DP}
\\
[K_m,M_{ns}]&=2\eta_{m[n}K_{s]}
\label{KM}
\\
[P_m,K_n]&=2\left(\eta_{mn}D+M_{mn}\right)
\label{PK}
\\
[D,K_m]&=-K_{m}.
\label{DK}
\end{align}
Here, the Latin indices run from $0$ to $3$ and serve to label Killing fields, e.g.~$M_{01}$ is one Killing field that can be written as a vector with five components $M_{01}^\mu$.

In \Bala\ it was claimed that the metrics \eqref{genAnsatz} and \eqref{specAnsatz} holographically describe a system in which conformal invariance (which includes continuous scale invariance) is broken down to \Poincare\ invariance in addition to discrete scale invariance. However, in \cite{Nakayama:2013is} the suspicion was raised that the spacetime \eqref{specAnsatz} may still admit the full algebra \eqref{MM}-\eqref{DK} of Killing fields in a somewhat hidden way. 

How can we settle this question? The general ansatz for a Killing field in a $5+1$ dimensional spacetime would be a vector field with six components which each might depend on all six coordinates. The Killing equations \eqref{Killing} would then lead to a set of coupled partial differential equations which in general will be prohibitively complicated and not allow for an easy solution. However, there are a few tricks that we can employ to try and solve the Killing equation analytically on the backgrounds \eqref{genAnsatz} and \eqref{specAnsatz}.\footnote{See \cite{Krtous:2015ona} for a general exploration of the question when Killing fields of one of the factor spacetimes will also generalise to Killing fields of a warped product spacetime, which we will however not make use of.} 

First of all, in \cite{Houri:2014hma,KYpackage,Houri:2015lxa}, a very useful algorithm was presented which can be used to place a (non-trivial) upper-bound on the number of Killing fields of a given spacetime. For \AdS, this algorithm yields the bound of $15$, which is expected as AdS is a maximally symmetric spacetime. For the metric \eqref{genAnsatz} with general functions\footnote{There is a possible source of confusion stemming from the use of the term \textit{upper-bound} here. The reader should be aware that when a metric depends on unspecified functions or parameters, the algorithm of \cite{Houri:2014hma,KYpackage,Houri:2015lxa} gives an upper bound for the number of linearly independent Killing fields for a generic choice for these functions or parameters. Upon fixing parameters or functions in a specific way, this upper bound may indeed increase.}  $A(w,\theta),B(w,\theta)$ and $C(w,\theta)$, the upper bound yielded is 10, confirming that this type of ansatz \textit{can} break the conformal algebra of continuous isometries down to the \Poincare\ algebra. For the metric \eqref{specAnsatz} and the choice $\alpha=0$, we obtain a bound of $16$, which is the expected result as explained at the end of section \ref{sec::Sol}. An interesting thing happens for the metric \eqref{specAnsatz} with $\alpha>0$, where we obtain an upper bound of 15. This indicates that the full conformal algebra \eqref{MM}-\eqref{DK} may indeed still be present, but of course does not constitute a proof of this suspicion. We will hence proceed to solve the Killing equation analytically on this background. 

\begin{figure}[htb]
	\centering
	\includegraphics[width=0.385\columnwidth]{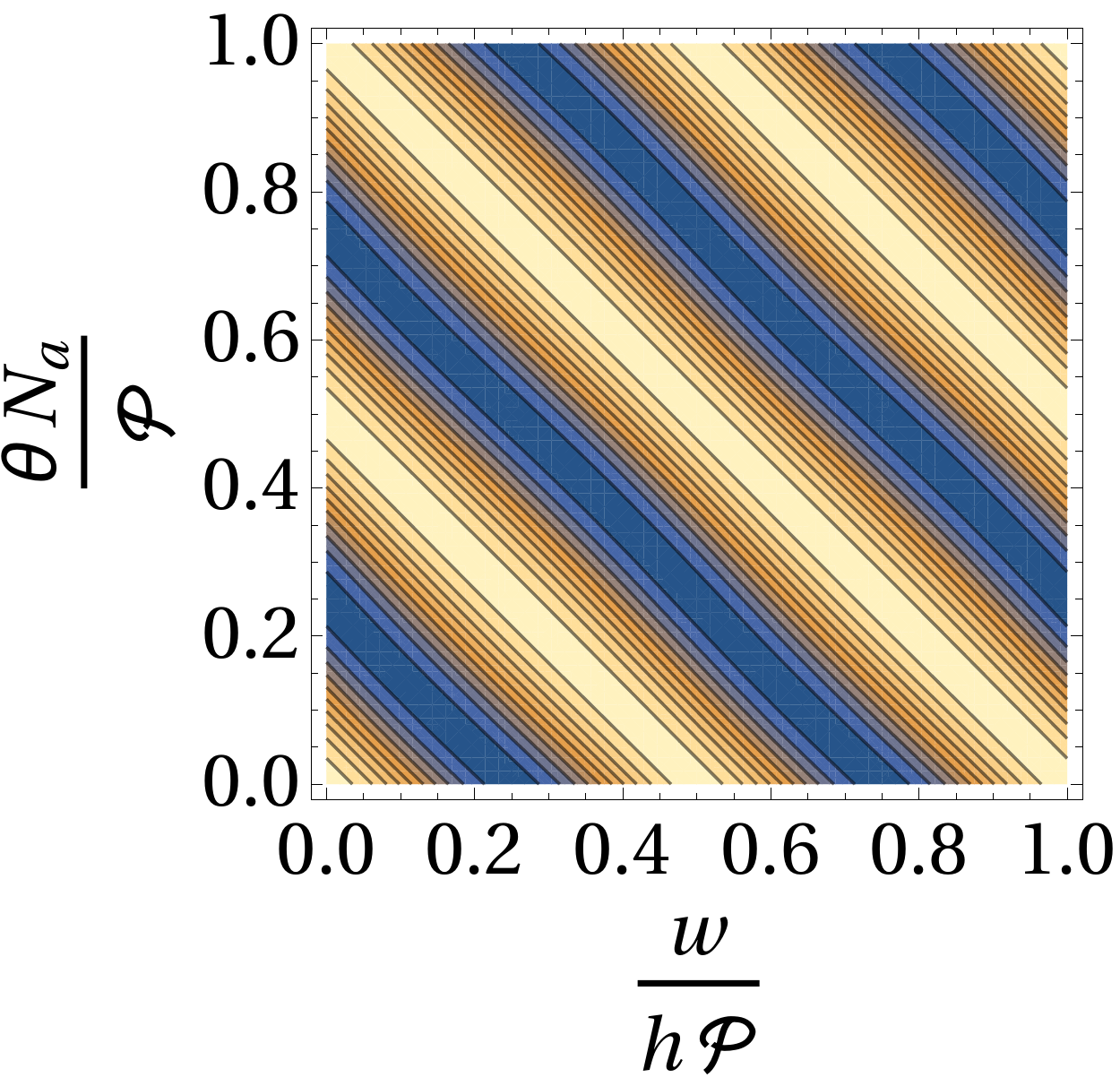}
	\includegraphics[width=0.1\columnwidth]{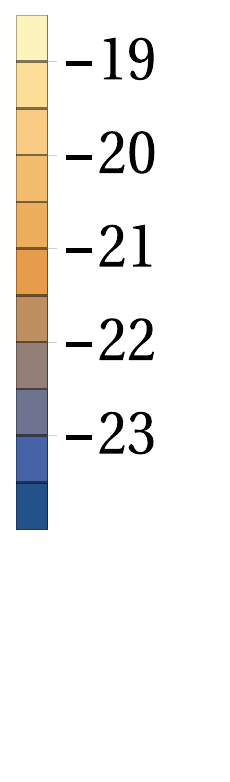}
	\includegraphics[width=0.385\columnwidth]{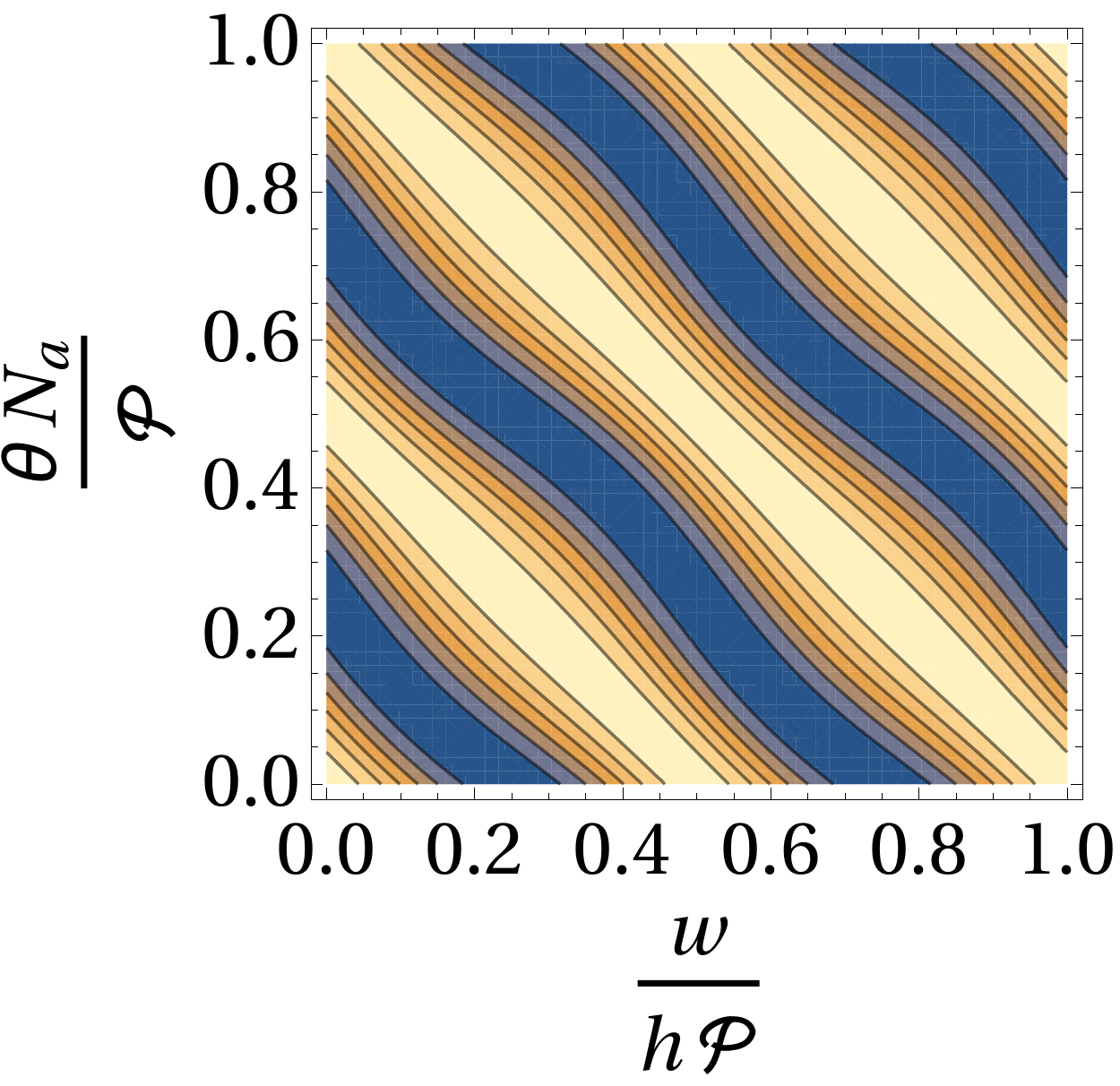}
	\includegraphics[width=0.1\columnwidth]{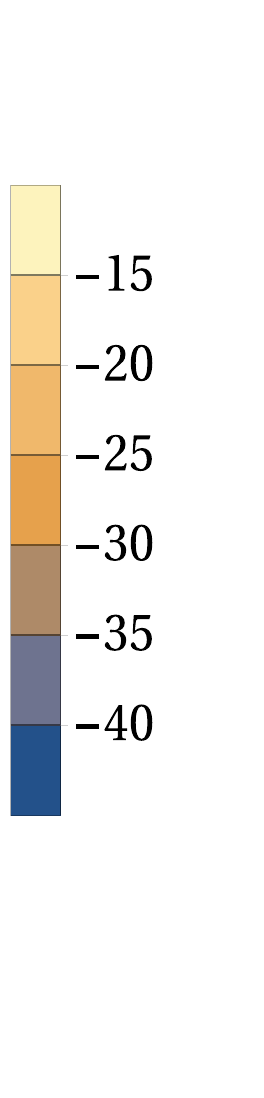}
	\caption[]{Contour-plots of the Ricci scalar $R(w,\theta)$ for $L=1$ and $\alpha=1/3$ (left) respectively $\alpha=2/3$ (right). $\mP$ is the $\alpha$-dependent period of the Jacobi elliptic functions \eqref{Period}. 
	}
	\label{fig::RS}
\end{figure}

As a Killing vector field describes the isometries of a spacetime, it is clear that when following the flow of such a vector field, physical properties of the spacetime such as curvature scalars should not change. This means that any Killing vector needs to satisfy the condition\footnote{By definition, Killing vector fields only describe the symmetries of the metric, it is however interesting to note that \eqref{RSconstraint} is equivalent to the condition $\mK^{\mu}\partial_{\mu}\chi=0$ with $\chi$ being given by \eqref{chi}. This means that in the solution \eqref{specAnsatz}, the matter fields that source the spacetime obey the same symmetries as the metric. This is in contrast to the models studied in \cite{Nakayama:2009qu,Nakayama:2010wx,Nakayama:2010ye,Nakayama:2011zw,Nakayama:2013is}, where the metric has a conformal symmetry group which is only broken by the presence of matter fields.}
\begin{align}
\mK^{\mu}\partial_{\mu}R=0,
\label{RSconstraint}
\end{align} 
where $R$ is the $w$ and $\theta$-dependent Ricci scalar of the metric \eqref{specAnsatz}. This yields a simple algebraic relation between the $w$ and $\theta$-components of any possible Killing field, which upon lowering the index implies the $\theta$-component to vanish for the background \eqref{specAnsatz}: $\mK_\theta=0$. See figure \ref{fig::RS} for contour-plots of the Ricci-scalar.

Furthermore, we can utilise the fact that on the spacetime background \eqref{specAnsatz}, we already know the 10 Killing vector fields $P_m$ and $M_{mn}$ which form the \Poincare\ algebra \eqref{MM}, \eqref{PM}. If continuous scale invariance is \textit{not} broken to its discrete counterpart, there should exist a Killing vector $D$ commuting with the $P_m$ and $M_{mn}$ according to \eqref{DP} and $[D,M_{mn}]=0$. As the Lie-bracket \eqref{bracket} does not depend on the specific metric or the connection, this condition allows us to constrain the generic form of a $D$-type Killing vector to
\begin{align}
D^{\mu}=\left(\begin{array}{c}
D^{t\ }\\ 
D^{x_1}\\ 
D^{x_2}\\ 
D^{x_3}\\ 
D^{w\ }\\ 
D^{\theta\ \ }
\end{array}  \right)=\left(\begin{array}{c}
-t \\ 
-x_1 \\ 
-x_2 \\ 
-x_3 \\ 
D^w(w,\theta)\\ 
D^\theta(w,\theta)
\end{array}  \right)
\label{Dgen}
\end{align}
for any metric of the form \eqref{genAnsatz}. As said above, the two components $D^w$ and $D^\theta$ can be related by equation \eqref{RSconstraint}. For the remaining component, say $D^w$, the $tt$-component of \eqref{Killing} then yields an algebraic equation. Furthermore, the $ww$, $w\theta$ and $\theta\theta$-components of \eqref{Killing} can be combined to yield another algebraic equation for $D^w$. This means that for a $D$-type Killing field on a spacetime \eqref{genAnsatz}, the ansatz \eqref{Dgen} leads to an overdetermined system of algebraic equations which in general will not have a solution, i.e.~continuous scale invariance will in general be broken. If it is not broken, however, the corresponding Killing field can be found by solving a set of purely algebraic equations for the components $D^w$ and $D^\theta$. In the case of the background metric \eqref{specAnsatz}, we indeed find the Killing field 
\begin{align}
D^{\mu}=\left(\begin{array}{c}
-t \\ 
-x_1 \\ 
-x_2 \\ 
-x_3 \\ 
L\\ 
-\frac{L \dn\left(w/h, \alpha^2\right)}{
h N_a  \dn\left(N_a \theta, \alpha^2\right)}
\end{array}  \right),
\label{D}
\end{align}

and similarly

\begin{align}
K_0^{\mu}=\left(\begin{array}{c}
e^{-2w/L} L^2 + t^2 + x_1^2 + x_2^2 + x_3^2 \\ 
2 t x_1 \\ 
2 t x_2 \\ 
2 t x_3 \\ 
-2 L t\\ 
\frac{2 L t \dn\left(w/h, \alpha^2\right)}{
h N_a \dn\left(N_a \theta, \alpha^2\right)}
\end{array}\right),
\label{K1}
\end{align}

\begin{align}
K_1^{\mu}=\left(\begin{array}{c}
-2 t x_1\\ 
e^{-2w/L} L^2 - t^2 - x_1^2 + x_2^2 + 
x_3^2 \\ 
-2 x_1 x_2 \\ 
-2 x_1 x_3 \\ 
2 L x_1\\ 
 -\frac{2 L x_1 \dn\left(w/h, \alpha^2\right)}{
h N_a \dn\left(N_a \theta, \alpha^2\right)}
\end{array}\right),
\label{K2}
\end{align}

\begin{align}
K_2^{\mu}=\left(\begin{array}{c}
-2 t x_2 \\ 
-2 x_1 x_2 \\ 
e^{-2w/L} L^2 - t^2 + x_1^2 - x_2^2 + x_3^2 \\ 
 -2 x_2 x_3 \\ 
2 L x_2\\ 
 -\frac{2 L x_2 \dn\left(w/h, \alpha^2\right)}{h N_a \dn\left(N_a \theta, \alpha^2\right)}
\end{array}\right),
\label{K3}
\end{align}

\begin{align}
K_3^{\mu}=\left(\begin{array}{c}
-2 t x_3\\ 
-2 x_1 x_3\\ 
-2 x_2 x_3\\ 
e^{-2w/L} L^2 - t^2 + x_1^2 + x_2^2 - x_3^2\\ 
2 L x_3\\ 
-\frac{2 L x_3 \dn\left(w/h, \alpha^2\right)}{
	h N_a \dn\left(N_a \theta, \alpha^2\right)}
\end{array}\right).
\label{K4}
\end{align}

This proves that the metric \eqref{specAnsatz} indeed retains the full conformal isometry group \eqref{MM}-\eqref{DK} of \AdS\ for any $\alpha$, instead of breaking it down to discrete scale invariance as one might naively have expected. As can be seen in equations \eqref{D}-\eqref{K4}, these Killing vector fields have non-zero components in the direction $\partial_\theta$ of the extra dimension. This is reminiscent of the holographic models for systems with Galilean or Schr\"odinger symmetry \cite{Son:2008ye,Balasubramanian:2008dm}.\footnote{Another important observation is that the Killing fields \eqref{D}-\eqref{K4} can be defined \textit{globally}. As said above, the $\theta$-direction is assumed to be identified periodically $\left(\theta\sim\theta+2\pi\right)$, so if the solutions to the Killing equation would have yielded vector fields with a prefactor $e^\theta$, for example, such a Killing field would only be defined locally, but globally it would not be possible to fix its norm. Such local Killing fields exist for example in the BTZ spacetime \cite{Banados:1992gq,Banados:1992wn} which is locally AdS (and hence posesses 6 local Killing fields), but only two Killing fields can be defined globally.} Based on the presence of this isometry group, we conjecture that on the side of the assumed holographically dual field theory, the cyclic RG flow can be shown to be equivalent to a conformal fixed point by a suitable field redefinition, similar to what was done in \cite{Luty:2012ww,Fortin:2012hn}. 

Of course it is possible to apply the same techniques to the $9+1$-dimensional top-down model that had also been studied in \Bala. Curiously, it is then easy (albeit tedious) to show that for the parameter $\alpha\neq0$\footnote{And under the assumption that the Killing vectors forming the isometry algebra \eqref{MM}-\eqref{DK} should commute with the left-over Killing vectors of the deformed $S^5$.}, this spacetime does \textit{not} support additional Killing vectors extending the \Poincare\ algebra to a conformal algebra, i.e.~there is \textit{no} hidden conformal symmetry present in the isometries of the top-down model of \Bala. It would certainly be interesting to have a closer look at the field theory side of this model.

\section{Scale without conformal invariance}
\label{sec::scale}

There have been attempts to find holographic models in which \Poincare\ invariance is combined with scale invariance but \textit{not} full conformal invariance, i.e.~in which only the algebra \eqref{MM}-\eqref{DP} holds, without the generators $K_m$ of special conformal transformations \cite{Nakayama:2009qu,Nakayama:2009fe,Nakayama:2010wx,Nakayama:2010ye,Nakayama:2011zw,Nakayama:2013is}\footnote{See also \cite{Awad:2000ac} for related work. Physical systems that exhibit scale without conformal invariance where discussed for example in \cite{Nakayama:2010ye,Jackiw:2011vz,Oz:2018yaz}}. Although in these papers the point has been made that this should not be possible in holography\footnote{See also \cite{Luty:2012ww,Fortin:2012hn} for field theory arguments.}, it would be interesting to find out whether an ansatz of the form \eqref{genAnsatz} can help to evade this no-go theorem, just like it helped to evade the no-go theorem for discrete scale invariance based on the holographic $c$-theorem in section \ref{sec::Intro}. This would lead to the question whether there can be physical solutions with a metric of the form \eqref{genAnsatz}, in which an additional $D$-type Killing vector \eqref{Dgen} exists, but no additional $K$-type Killing vectors. In section \ref{sec::Killing}, we have seen that on a spacetime with metric \eqref{genAnsatz}, any Killing vector $D$ that fits together with the \Poincare\ algebra according to \eqref{MM}-\eqref{DP} has to have the form \eqref{Dgen}. In general, there will be no such Killing vector, but let us make the assumption that a Killing vector of the form \eqref{Dgen} does exist on some metric of the form \eqref{genAnsatz}. Does this assumption \textit{imply} the presence of four additional Killing vectors $K_m$? Based on the algebra \eqref{MM}-\eqref{DK}, such vector fields would necessarily have the form
\begin{align}
K_0^{\mu}&=\left(\begin{array}{c}
\mK(w,\theta) + t^2 + x_1^2 + x_2^2 + x_3^2 \\ 
2 t x_1 \\ 
2 t x_2 \\ 
2 t x_3 \\ 
-2  t D^w(w,\theta)\\ 
-2  t D^\theta(w,\theta)
\end{array}\right)
\label{K1gen}
\end{align}
with the other $K_m$ similarly, and with the function $\mK(w,\theta)$ satisfying the equation
\begin{align}
2 \mK(w,\theta) +D^\theta(w,\theta)\partial_\theta\mK(w,\theta)+D^w(w,\theta)\partial_w\mK(w,\theta)=0.
\label{commutator}
\end{align}
The assumption that $D$ (equation \eqref{Dgen}) is a Killing vector directly implies that also several components of the Killing equation for $K_0$ will vanish, so that we are left with
\begin{align}
0=&2A(w,\theta )e^{2B(w,\theta)}
\left(A(w,\theta) D^w(w,\theta)+D^\theta(w,\theta)\right)
\nonumber
\\
&
+e^{2C(w,\theta)} \left(e^{\frac{2 w}{L}}\partial_w \mK(w,\theta)+2D^w(w,\theta)\right),
\label{Killing15}
\\
0=&2 e^{2B(w,\theta)} (A(w,\theta ) D^w(w,\theta)+D^\theta(w,\theta))
+e^{\frac{2 w}{L}} e^{2C(w,\theta)} \partial_\theta\mK(w,\theta).
\label{Killing16}
\end{align}
Again, this deceptively simple set of equations \eqref{commutator}-\eqref{Killing16} for the function $\mK(w,\theta)$ seems to be overdetermined: There may not be a solution in general, but if there is a solution, it can be found by combining the equations into an algebraic equation for $\mK(w,\theta)$, and solving it. The mere existence of a Killing vector $D$ (equation \eqref{Dgen}) does not appear in any obvious way to be sufficient to infer that $K_0$ (equation \eqref{K1gen}) is also automatically a Killing vector (i.e.~that equations \eqref{commutator}-\eqref{Killing16} have a solution). It would be interesting to find out precisely under which assumptions the one implies the other. We will leave these interesting questions for future research. 

\section{Discussion and Outlook}
\label{sec::conc}

What did we learn in this paper? \Bala\ set out to formulate a bottom-up holographic model of a discretely scale invariant \Poincare\ invariant field theory. Generically, one would expect such a model to satisfy the following properties:
\begin{itemize}
	\item The isometry group of the metric should combine \Poincare\ invariance with discrete scale invariance.
	\item The metric should be sourced, via Einstein's equations, by a matter distribution which satisfies the NEC.
	\item The system should be (at the very least perturbatively) stable.
\end{itemize}
It was shown in \Bala\ that the solution \eqref{specAnsatz} satisfies these requirements. In addition to this, the following statements hold:
\begin{itemize}
	\item The metric is sourced, via Einstein's equations, by only one real scalar field (equation \eqref{EinsteinEq}) \Bala.
	\item The Weyl tensor vanishes identically.
\end{itemize}
Furthermore, it was the main point of this paper to prove that the following is true:
\begin{itemize}	
	\item The system exhibits a hidden conformal isometry group \eqref{MM}-\eqref{DK} due to the existence of the Killing vector fields \eqref{D}-\eqref{K4}.\footnote{A search for higher Killing-Yano-forms along the lines of \cite{Houri:2014hma,KYpackage,Houri:2015lxa} is left for the future. } 
\end{itemize}
It would be interesting to find out whether this last point follows mathematically from one or more of the other points, or whether it is possible to construct a holographic bottom-up model satisfying at least the first three requirements without a hidden conformal symmetry. We will leave this as an intriguing problem for future research.  

There is an important questions of recent interest to the holography community that appears to be related to the quest for a holographic model of discretely scale invariant systems: As discussed above the compact extra dimension introduced in the ansatz \eqref{genAnsatz} is stabilised by the flux of the axion-like scalar field $\chi$ in the solution \eqref{specAnsatz}. However, in the non-supersymmetric case, the possible stability of such spacetimes has been called into question recently \cite{Ooguri:2016pdq}. Hence, in investigations of bulk models that (may) allow for cyclic RG flows, the issue of stability will be of particular relevance. It may be worthwhile to revisit the top-down model of \Bala\  in this context.

So far we have focused on the combination of  discrete scale invariance with \Poincare-invariance. This is an interesting research direction, in that it could yield insights into what kind of RG flows can in principle be studied holographically. There are four scenarios concerning the admissibility of discrete scale invariance in holography:

Firstly, it could be that field theory arguments, such as the $a$-theorem \cite{Komargodski:2011vj}, the $c$-theorem \cite{Zamolodchikov:1986gt}, or other stronger ones still to be discovered, imply that discrete scale invariance is generically not allowed on the field theory side (for a given number of dimensions). One would then expect that a similar no-go theorem has to exist on the gravity side of AdS/CFT models. In mathematical physics, the importance of a theorem sometimes turns out not to lie entirely in the statement of the theorem, but also in the development of the techniques which where necessary to complete the proof.  Our intuition hereby is that a particularly important or fundamental theorem on the field theory side might correspond to a result which is equally impactful on the gravity side of the holographic duality. Seeking a proof of this statement in terms of gravitational physics may then yield insights into the nature of holographic duality that go beyond the mere statement of the theorem itself. 

Secondly, there is the logical possibility that some kind of no-go theorem exists on the boundary side which has no counterpart on the gravity side of AdS/CFT.  This would mean that bulk models exhibiting discrete scale invariance are possible, but cannot have a well defined holographic dual. These bulk models would hence belong to a "forbidden landscape" or "swampland" of holography, a view taken in \cite{Nakayama:2009qu,Nakayama:2009fe}.

The third possibility is that discrete scale invariance is allowed both on the field theory side and the bulk side of holography. Then, holography could be used as a model building tool to study specific properties of cyclic RG flows, similar to the way holography has been used to study other types of RG flows in \cite{Freedman:1999gp,Franco:2004jz,Shaghoulian:2016umj,Donos:2017ljs,Donos:2017sba,Kiritsis:2016kog}  and many other works.

Fourthly, it could be that discrete scale invariance is forbidden on the gravity side, but allowable in field theories. This would imply that the emergence of discrete scale invariance  can be used as a diagnostic tool to find out which field theories can have a gravity dual in principle and which ones can not. This is similar to the way in which certain entanglement inequalities which can be proven holographically can be used \cite{Hayden:2011ag}. Restricting to field theories with a holographic dual, this case however reduces to the first possibility discussed above. 

In any of these cases, we believe that the study of discrete scale invariance in holography might yield interesting new results. Especially the top-down model of \Bala\  may be a worthwhile research subject in this context, as the string theory construction provides a good understanding of the dual field theory. Questions of stability, geodesic completeness, or absence of singularities in this model are however left for future research.

Beyond questions of mathematical proofs of concept, where do we expect discrete scale invariance (with or without additional \Poincare\ invariance) to play a role in physics? 

In the introduction, we have given a number of references concerning the appearance of discrete scale invariance in physics \cite{Sornette:1997pb,SORNETTE1997411,refId0,refId1,0295-5075-41-1-043,saleur:hal-00331040,Choptuik:1992jv,Abrahams:1993wa,Hirschmann:1994du,Hirschmann:1995qx,Gralla:2017lto,Kol:2002xz,Kalisch:2017bin,Hammer:2011kg,Georgi:2016tjs,Calcagni:2017via,0305-4470-29-13-017,PhysRevB.90.184202,Wilson:1970ag,Glazek:2002hq,LeClair:2002ux,LeClair:2003hj,Curtright:2011qg,Bulycheva:2014twa}, however only \cite{Liu:2009dm,Faulkner:2009wj,Hartnoll:2015rza,Erdmenger:2016msd,Brattan:2017yzx,Balasubramanian:2013ux,Nakayama:2011zw,Nakayama:2013is} where in a holographic context. 

We would however also like to point out that tensor networks of the MERA type \cite{Vidal:2007hda} by definition only implement a discrete version of scale invariance. When comparing the states described by such MERA networks to holographic systems, it is often assumed that the appropriate metric arising from the MERA network in some continuum limit is the metric of AdS or dS space  (see \cite{Swingle:2009bg,Czech:2015kbp}), i.e.~that a continuous form of scale invariance (paired with full conformal invariance) arises from the underlying tensor network, similarly to how a continuous translation invariance can appear from a discrete lattice. It might be useful to study whether this is always true, or whether the underlying discrete scale invariance can still leave some traces after the continuum limit.    

Going even further, we might wonder whether it is possible to force a field theory or condensed matter model to exhibit discrete scale invariance by artificially placing it on a fractal lattice, see \cite{PhysRevLett.45.855} for early ideas in this direction. Given the possibilities of Nanotechnology (e.g.~to construct quantum dots and nanowires), one might even envision that it could be possible to build a system in the laboratory which lives on a space that is fractal at least over a certain range of scales.

On a more abstract level, the topic of this paper was how to find spacetimes with particular desired symmetry properties out of a large family of models (in our case metrics of the form  \eqref{genAnsatz}). It might be interesting to study in a similar way other sufficiently diverse families of models with a holographic interpretation, such as the BKL like spacetimes of \cite{Shaghoulian:2016umj,Ren:2016xhb} or the LLM geometries of \cite{Lin:2004nb}\footnote{The latter have, in their ten-dimensional form, at least an $SO(4)\times SO(4)\times \mathbb{R}$ isometry group, but of course specific examples with more isometries exist, $AdS_5\times S^5$ being the most obvious example}, in search of examples with unexpected or unusual symmetry properties.

\section*{Acknowledgements}

The author would like to thank Martin Ammon, Johanna Erdmenger, Jerome Gauntlett, Sam Gralla, Romuald Janik, Ren\'{e} Meyer, Nina Miekley, Max-Niklas Newrzella, Charles Melby-Thompson, and \'Alvaro V\'eliz-Osorio for useful discussions, as well as Koushik Balasubramanian for useful discussions and for sharing some of his computer code. The author was supported by the Polish National Science Centre (NCN) grant 2012/06/A/ST2/00396 until the 16th.~of November 2017, and the NCN grant 2017/24/C/ST2/00469 starting from the 17th.~of November 2017. Some of the calculations for this work were done using the \textit{Mathematica} package \texttt{KY\_upperbound} \cite{KYpackage}.


\begin{thebibliography}{10}

\bibitem{Balasubramanian:2013ux}
K.~Balasubramanian, ``{Gravity duals of cyclic RG flows, with strings
  attached},''
\href{http://arxiv.org/abs/1301.6653}{{\ttfamily arXiv:1301.6653 [hep-th]}}.

\bibitem{Sornette:1997pb}
D.~Sornette, ``{Discrete scale invariance and complex dimensions},''
  \href{http://dx.doi.org/10.1016/S0370-1573(97)00076-8}{{\em Phys. Rept.}
  {\bfseries 297} (1998) 239--270},
\href{http://arxiv.org/abs/cond-mat/9707012}{{\ttfamily arXiv:cond-mat/9707012
  [cond-mat.stat-mech]}}.

\bibitem{SORNETTE1997411}
D.~Sornette and A.~Johansen, ``Large financial crashes,''
  \href{http://dx.doi.org/https://doi.org/10.1016/S0378-4371(97)00318-X}{{\em
  Physica A: Statistical Mechanics and its Applications} {\bfseries 245} no.~3,
  (1997) 411 -- 422}.
  \url{http://www.sciencedirect.com/science/article/pii/S037843719700318X}.

\bibitem{refId0}
{Didier Sornette}, {Anders Johansen}, and {Jean-Philippe Bouchaud}, ``Stock
  market crashes, precursors and replicas,''
  \href{http://dx.doi.org/10.1051/jp1:1996135}{{\em J. Phys. I France}
  {\bfseries 6} no.~1, (1996) 167--175}.
  \url{https://doi.org/10.1051/jp1:1996135}.

\bibitem{refId1}
{Didier Sornette} and {Charles G. Sammis}, ``Complex critical exponents from
  renormalization group theory of earthquakes: Implications for earthquake
  predictions,'' \href{http://dx.doi.org/10.1051/jp1:1995154}{{\em J. Phys. I
  France} {\bfseries 5} no.~5, (1995) 607--619}.
  \url{https://doi.org/10.1051/jp1:1995154}.

\bibitem{0295-5075-41-1-043}
Y.~Huang, H.~Saleur, C.~Sammis, and D.~Sornette, ``Precursors, aftershocks,
  criticality and self-organized criticality,'' {\em EPL (Europhysics Letters)}
  {\bfseries 41} no.~1, (1998) 43.
  \url{http://stacks.iop.org/0295-5075/41/i=1/a=043}.

\bibitem{saleur:hal-00331040}
H.~Saleur, C.~G. Sammis, and D.~Sornette, ``{Renormalization group theory of
  earthquakes},'' {\em {Nonlinear Processes in Geophysics}} {\bfseries 3}
  no.~2, (1996) 102--109. \url{https://hal.archives-ouvertes.fr/hal-00331040}.

\bibitem{Choptuik:1992jv}
M.~W. Choptuik, ``{Universality and scaling in gravitational collapse of a
  massless scalar field},''
\href{http://dx.doi.org/10.1103/PhysRevLett.70.9}{{\em Phys. Rev. Lett.}
  {\bfseries 70} (1993) 9--12}.

\bibitem{Abrahams:1993wa}
A.~M. Abrahams and C.~R. Evans, ``{Critical behavior and scaling in vacuum
  axisymmetric gravitational collapse},''
\href{http://dx.doi.org/10.1103/PhysRevLett.70.2980}{{\em Phys. Rev. Lett.}
  {\bfseries 70} (1993) 2980--2983}.

\bibitem{Hirschmann:1994du}
E.~W. Hirschmann and D.~M. Eardley, ``{Universal scaling and echoing in
  gravitational collapse of a complex scalar field},''
  \href{http://dx.doi.org/10.1103/PhysRevD.51.4198}{{\em Phys. Rev.} {\bfseries
  D51} (1995) 4198--4207},
\href{http://arxiv.org/abs/gr-qc/9412066}{{\ttfamily arXiv:gr-qc/9412066
  [gr-qc]}}.

\bibitem{Hirschmann:1995qx}
E.~W. Hirschmann and D.~M. Eardley, ``{Critical exponents and stability at the
  black hole threshold for a complex scalar field},''
  \href{http://dx.doi.org/10.1103/PhysRevD.52.5850}{{\em Phys. Rev.} {\bfseries
  D52} (1995) 5850--5856},
\href{http://arxiv.org/abs/gr-qc/9506078}{{\ttfamily arXiv:gr-qc/9506078
  [gr-qc]}}.

\bibitem{Gralla:2017lto}
S.~E. Gralla and P.~Zimmerman, ``{Critical Exponents of Extremal Kerr
  Perturbations},''
\href{http://arxiv.org/abs/1711.00855}{{\ttfamily arXiv:1711.00855 [gr-qc]}}.

\bibitem{Kol:2002xz}
B.~Kol, ``{Topology change in general relativity, and the black hole black
  string transition},''
  \href{http://dx.doi.org/10.1088/1126-6708/2005/10/049}{{\em JHEP} {\bfseries
  10} (2005) 049},
\href{http://arxiv.org/abs/hep-th/0206220}{{\ttfamily arXiv:hep-th/0206220
  [hep-th]}}.

\bibitem{Kalisch:2017bin}
M.~Kalisch, S.~M{\"o}ckel, and M.~Ammon, ``{Critical behavior of the black
  hole/black string transition},''
  \href{http://dx.doi.org/10.1007/JHEP08(2017)049}{{\em JHEP} {\bfseries 08}
  (2017) 049},
\href{http://arxiv.org/abs/1706.02323}{{\ttfamily arXiv:1706.02323 [gr-qc]}}.

\bibitem{Hammer:2011kg}
H.-W. Hammer and L.~Platter, ``{Efimov physics from a renormalization group
  perspective},'' \href{http://dx.doi.org/10.1098/rsta.2011.0001}{{\em Phil.
  Trans. Roy. Soc. Lond.} {\bfseries A369} (2011) 2679},
\href{http://arxiv.org/abs/1102.3789}{{\ttfamily arXiv:1102.3789 [nucl-th]}}.

\bibitem{Georgi:2016tjs}
H.~Georgi, ``{Physics Fun with Discrete Scale Invariance},''
\href{http://arxiv.org/abs/1606.03405}{{\ttfamily arXiv:1606.03405 [hep-ph]}}.

\bibitem{Calcagni:2017via}
G.~Calcagni, ``{Complex dimensions and their observability},''
  \href{http://dx.doi.org/10.1103/PhysRevD.96.046001}{{\em Phys. Rev.}
  {\bfseries D96} no.~4, (2017) 046001},
\href{http://arxiv.org/abs/1705.01619}{{\ttfamily arXiv:1705.01619 [gr-qc]}}.

\bibitem{0305-4470-29-13-017}
D.~Karevski and L.~Turban, ``Log-periodic corrections to scaling: exact results
  for aperiodic ising quantum chains,'' {\em Journal of Physics A: Mathematical
  and General} {\bfseries 29} no.~13, (1996) 3461.
  \url{http://stacks.iop.org/0305-4470/29/i=13/a=017}.

\bibitem{PhysRevB.90.184202}
G.~m.~H. Ro\'osz, U.~Divakaran, H.~Rieger, and F.~Igl\'oi, ``Nonequilibrium
  quantum relaxation across a localization-delocalization transition,''
  \href{http://dx.doi.org/10.1103/PhysRevB.90.184202}{{\em Phys. Rev. B}
  {\bfseries 90} (Nov, 2014) 184202}.
  \url{https://link.aps.org/doi/10.1103/PhysRevB.90.184202}.

\bibitem{Liu:2009dm}
H.~Liu, J.~McGreevy, and D.~Vegh, ``{Non-Fermi liquids from holography},''
  \href{http://dx.doi.org/10.1103/PhysRevD.83.065029}{{\em Phys. Rev.}
  {\bfseries D83} (2011) 065029},
\href{http://arxiv.org/abs/0903.2477}{{\ttfamily arXiv:0903.2477 [hep-th]}}.

\bibitem{Faulkner:2009wj}
T.~Faulkner, H.~Liu, J.~McGreevy, and D.~Vegh, ``{Emergent quantum criticality,
  Fermi surfaces, and AdS(2)},''
  \href{http://dx.doi.org/10.1103/PhysRevD.83.125002}{{\em Phys. Rev.}
  {\bfseries D83} (2011) 125002},
\href{http://arxiv.org/abs/0907.2694}{{\ttfamily arXiv:0907.2694 [hep-th]}}.

\bibitem{Hartnoll:2015rza}
S.~A. Hartnoll, D.~M. Ramirez, and J.~E. Santos, ``{Thermal conductivity at a
  disordered quantum critical point},''
  \href{http://dx.doi.org/10.1007/JHEP04(2016)022}{{\em JHEP} {\bfseries 04}
  (2016) 022},
\href{http://arxiv.org/abs/1508.04435}{{\ttfamily arXiv:1508.04435 [hep-th]}}.

\bibitem{Erdmenger:2016msd}
J.~Erdmenger, M.~Flory, M.-N. Newrzella, M.~Strydom, and J.~M.~S. Wu,
  ``{Quantum Quenches in a Holographic Kondo Model},''
  \href{http://dx.doi.org/10.1007/JHEP04(2017)045}{{\em JHEP} {\bfseries 04}
  (2017) 045},
\href{http://arxiv.org/abs/1612.06860}{{\ttfamily arXiv:1612.06860 [hep-th]}}.

\bibitem{Brattan:2017yzx}
D.~K. Brattan, O.~Ovdat, and E.~Akkermans, ``{Scale anomaly of a Lifshitz
  scalar: a universal quantum phase transition to discrete scale invariance},''
\href{http://arxiv.org/abs/1706.00016}{{\ttfamily arXiv:1706.00016 [hep-th]}}.

\bibitem{Wilson:1970ag}
K.~G. Wilson, ``{The Renormalization Group and Strong Interactions},''
\href{http://dx.doi.org/10.1103/PhysRevD.3.1818}{{\em Phys. Rev.} {\bfseries
  D3} (1971) 1818}.

\bibitem{Glazek:2002hq}
S.~D. Glazek and K.~G. Wilson, ``{Limit cycles in quantum theories},''
  \href{http://dx.doi.org/10.1103/PhysRevLett.89.230401}{{\em Phys. Rev. Lett.}
  {\bfseries 89} (2002) 230401},
  \href{http://arxiv.org/abs/hep-th/0203088}{{\ttfamily arXiv:hep-th/0203088
  [hep-th]}}.
[Erratum: Phys. Rev. Lett.92,139901(2004)].

\bibitem{LeClair:2002ux}
A.~LeClair, J.~M. Roman, and G.~Sierra, ``{Russian doll renormalization group
  and superconductivity},''
  \href{http://dx.doi.org/10.1103/PhysRevB.69.020505}{{\em Phys. Rev.}
  {\bfseries B69} (2004) 020505},
\href{http://arxiv.org/abs/cond-mat/0211338}{{\ttfamily arXiv:cond-mat/0211338
  [cond-mat]}}.

\bibitem{LeClair:2003hj}
A.~LeClair, J.~M. Roman, and G.~Sierra, ``{Log periodic behavior of finite size
  effects in field theories with RG limit cycles},''
  \href{http://dx.doi.org/10.1016/j.nuclphysb.2004.08.033}{{\em Nucl. Phys.}
  {\bfseries B700} (2004) 407--435},
\href{http://arxiv.org/abs/hep-th/0312141}{{\ttfamily arXiv:hep-th/0312141
  [hep-th]}}.

\bibitem{Curtright:2011qg}
T.~L. Curtright, X.~Jin, and C.~K. Zachos, ``{RG flows, cycles, and c-theorem
  folklore},'' \href{http://dx.doi.org/10.1103/PhysRevLett.108.131601}{{\em
  Phys. Rev. Lett.} {\bfseries 108} (2012) 131601},
\href{http://arxiv.org/abs/1111.2649}{{\ttfamily arXiv:1111.2649 [hep-th]}}.

\bibitem{Bulycheva:2014twa}
K.~M. Bulycheva and A.~S. Gorsky, ``{Limit cycles in renormalization group
  dynamics},'' \href{http://dx.doi.org/10.3367/UFNe.0184.201402g.0182}{{\em
  Phys. Usp.} {\bfseries 57} (2014) 171--182},
  \href{http://arxiv.org/abs/1402.2431}{{\ttfamily arXiv:1402.2431 [hep-th]}}.
[Usp. Fiz. Nauk184,no.2,182(2014)].

\bibitem{Luty:2012ww}
M.~A. Luty, J.~Polchinski, and R.~Rattazzi, ``{The $a$-theorem and the
  Asymptotics of 4D Quantum Field Theory},''
  \href{http://dx.doi.org/10.1007/JHEP01(2013)152}{{\em JHEP} {\bfseries 01}
  (2013) 152},
\href{http://arxiv.org/abs/1204.5221}{{\ttfamily arXiv:1204.5221 [hep-th]}}.

\bibitem{Fortin:2012hn}
J.-F. Fortin, B.~Grinstein, and A.~Stergiou, ``{Limit Cycles and Conformal
  Invariance},'' \href{http://dx.doi.org/10.1007/JHEP01(2013)184}{{\em JHEP}
  {\bfseries 01} (2013) 184},
\href{http://arxiv.org/abs/1208.3674}{{\ttfamily arXiv:1208.3674 [hep-th]}}.

\bibitem{Franco:2004jz}
S.~Franco, Y.-H. He, C.~Herzog, and J.~Walcher, ``{Chaotic duality in string
  theory},'' \href{http://dx.doi.org/10.1103/PhysRevD.70.046006}{{\em Phys.
  Rev.} {\bfseries D70} (2004) 046006},
\href{http://arxiv.org/abs/hep-th/0402120}{{\ttfamily arXiv:hep-th/0402120
  [hep-th]}}.

\bibitem{Shaghoulian:2016umj}
E.~Shaghoulian and H.~Wang, ``{Timelike BKL singularities and chaos in
  AdS/CFT},'' \href{http://dx.doi.org/10.1088/0264-9381/33/12/125020}{{\em
  Class. Quant. Grav.} {\bfseries 33} no.~12, (2016) 125020},
\href{http://arxiv.org/abs/1601.02599}{{\ttfamily arXiv:1601.02599 [hep-th]}}.

\bibitem{Ammon:2015wua}
M.~Ammon and J.~Erdmenger, {\em {Gauge/gravity duality}}.
\newblock Cambridge Univ. Pr., Cambridge, UK, 2015.

\bibitem{Freedman:1999gp}
D.~Z. Freedman, S.~S. Gubser, K.~Pilch, and N.~P. Warner, ``{Renormalization
  group flows from holography supersymmetry and a c theorem},'' {\em Adv.
  Theor. Math. Phys.} {\bfseries 3} (1999) 363--417,
\href{http://arxiv.org/abs/hep-th/9904017}{{\ttfamily arXiv:hep-th/9904017
  [hep-th]}}.

\bibitem{Donos:2017ljs}
A.~Donos, J.~P. Gauntlett, C.~Rosen, and O.~Sosa-Rodriguez, ``{Boomerang RG
  flows in M-theory with intermediate scaling},''
  \href{http://dx.doi.org/10.1007/JHEP07(2017)128}{{\em JHEP} {\bfseries 07}
  (2017) 128},
\href{http://arxiv.org/abs/1705.03000}{{\ttfamily arXiv:1705.03000 [hep-th]}}.

\bibitem{Donos:2017sba}
A.~Donos, J.~P. Gauntlett, C.~Rosen, and O.~Sosa-Rodriguez, ``{Boomerang RG
  flows with intermediate conformal invariance},''
\href{http://arxiv.org/abs/1712.08017}{{\ttfamily arXiv:1712.08017 [hep-th]}}.

\bibitem{Kiritsis:2016kog}
E.~Kiritsis, F.~Nitti, and L.~Silva~Pimenta, ``{Exotic RG Flows from
  Holography},'' \href{http://dx.doi.org/10.1002/prop.201600120}{{\em Fortsch.
  Phys.} {\bfseries 65} no.~2, (2017) 1600120},
\href{http://arxiv.org/abs/1611.05493}{{\ttfamily arXiv:1611.05493 [hep-th]}}.

\bibitem{Curiel:2014zba}
E.~Curiel, ``{A Primer on Energy Conditions},''
\href{http://arxiv.org/abs/1405.0403}{{\ttfamily arXiv:1405.0403
  [physics.hist-ph]}}.

\bibitem{Nakayama:2011zw}
Y.~Nakayama, ``{Gravity Dual for Cyclic Renormalization Group Flow without
  Conformal Invariance},''
  \href{http://dx.doi.org/10.1142/S0217732311036930}{{\em Mod. Phys. Lett.}
  {\bfseries A26} (2011) 2469--2476},
\href{http://arxiv.org/abs/1107.2928}{{\ttfamily arXiv:1107.2928 [hep-th]}}.

\bibitem{Nakayama:2013is}
Y.~Nakayama, ``{Scale invariance vs conformal invariance},''
  \href{http://dx.doi.org/10.1016/j.physrep.2014.12.003}{{\em Phys. Rept.}
  {\bfseries 569} (2015) 1--93},
\href{http://arxiv.org/abs/1302.0884}{{\ttfamily arXiv:1302.0884 [hep-th]}}.

\bibitem{Zamolodchikov:1986gt}
A.~B. Zamolodchikov, ``{Irreversibility of the Flux of the Renormalization
  Group in a 2D Field Theory},'' {\em JETP Lett.} {\bfseries 43} (1986)
  730--732.
[Pisma Zh. Eksp. Teor. Fiz.43,565(1986)].

\bibitem{Komargodski:2011vj}
Z.~Komargodski and A.~Schwimmer, ``{On Renormalization Group Flows in Four
  Dimensions},'' \href{http://dx.doi.org/10.1007/JHEP12(2011)099}{{\em JHEP}
  {\bfseries 12} (2011) 099},
\href{http://arxiv.org/abs/1107.3987}{{\ttfamily arXiv:1107.3987 [hep-th]}}.

\bibitem{Nakayama:2009qu}
Y.~Nakayama, ``{Forbidden Landscape from Holography},''
  \href{http://dx.doi.org/10.1088/1126-6708/2009/11/061}{{\em JHEP} {\bfseries
  11} (2009) 061},
\href{http://arxiv.org/abs/0907.0227}{{\ttfamily arXiv:0907.0227 [hep-th]}}.

\bibitem{Nakayama:2009fe}
Y.~Nakayama, ``{No Forbidden Landscape in String/M-theory},''
  \href{http://dx.doi.org/10.1007/JHEP01(2010)030}{{\em JHEP} {\bfseries 01}
  (2010) 030},
\href{http://arxiv.org/abs/0909.4297}{{\ttfamily arXiv:0909.4297 [hep-th]}}.

\bibitem{Freedman:2012zz}
D.~Z. Freedman and A.~Van~Proeyen, {\em {Supergravity}}.
\newblock Cambridge Univ. Press, Cambridge, UK, 2012.

\bibitem{Krtous:2015ona}
P.~Krtous, D.~Kubiznak, and I.~Kolar, ``{Killing-Yano forms and Killing tensors
  on a warped space},''
  \href{http://dx.doi.org/10.1103/PhysRevD.93.024057}{{\em Phys. Rev.}
  {\bfseries D93} no.~2, (2016) 024057},
\href{http://arxiv.org/abs/1508.02642}{{\ttfamily arXiv:1508.02642 [gr-qc]}}.

\bibitem{Houri:2014hma}
T.~Houri and Y.~Yasui, ``{A simple test for spacetime symmetry},''
  \href{http://dx.doi.org/10.1088/0264-9381/32/5/055002}{{\em Class. Quant.
  Grav.} {\bfseries 32} no.~5, (2015) 055002},
\href{http://arxiv.org/abs/1410.1023}{{\ttfamily arXiv:1410.1023 [gr-qc]}}.

\bibitem{KYpackage}
T.~Houri and Y.~Yasui, ``{KY$\_$upperbound \textit{Mathematica} package}.''
  \url{http://www.research.kobe-u.ac.jp/fsci-pacos/KY_upperbound/}, v1.2; 13th
  August, 2014.

\bibitem{Houri:2015lxa}
T.~Houri, ``{An upper bound on the number of Killing-Yano tensors},''
\href{http://dx.doi.org/10.1088/1742-6596/600/1/012064}{{\em J. Phys. Conf.
  Ser.} {\bfseries 600} no.~1, (2015) 012064}.

\bibitem{Nakayama:2010wx}
Y.~Nakayama, ``{Higher derivative corrections in holographic
  Zamolodchikov-Polchinski theorem},''
  \href{http://dx.doi.org/10.1140/epjc/s10052-012-1870-z}{{\em Eur. Phys. J.}
  {\bfseries C72} (2012) 1870},
\href{http://arxiv.org/abs/1009.0491}{{\ttfamily arXiv:1009.0491 [hep-th]}}.

\bibitem{Nakayama:2010ye}
Y.~Nakayama, ``{Gravity Dual for a Model of Perception},''
  \href{http://dx.doi.org/10.1016/j.aop.2010.09.009}{{\em Annals Phys.}
  {\bfseries 326} (2011) 2--14},
\href{http://arxiv.org/abs/1003.5729}{{\ttfamily arXiv:1003.5729 [hep-th]}}.

\bibitem{Son:2008ye}
D.~T. Son, ``{Toward an AdS/cold atoms correspondence: A Geometric realization
  of the Schrodinger symmetry},''
  \href{http://dx.doi.org/10.1103/PhysRevD.78.046003}{{\em Phys. Rev.}
  {\bfseries D78} (2008) 046003},
\href{http://arxiv.org/abs/0804.3972}{{\ttfamily arXiv:0804.3972 [hep-th]}}.

\bibitem{Balasubramanian:2008dm}
K.~Balasubramanian and J.~McGreevy, ``{Gravity duals for non-relativistic
  CFTs},'' \href{http://dx.doi.org/10.1103/PhysRevLett.101.061601}{{\em Phys.
  Rev. Lett.} {\bfseries 101} (2008) 061601},
\href{http://arxiv.org/abs/0804.4053}{{\ttfamily arXiv:0804.4053 [hep-th]}}.

\bibitem{Banados:1992gq}
M.~Banados, M.~Henneaux, C.~Teitelboim, and J.~Zanelli, ``{Geometry of the
  (2+1) black hole},'' \href{http://dx.doi.org/10.1103/PhysRevD.48.1506}{{\em
  Phys. Rev.} {\bfseries D48} (1993) 1506--1525},
  \href{http://arxiv.org/abs/gr-qc/9302012}{{\ttfamily arXiv:gr-qc/9302012
  [gr-qc]}}.
[Erratum: Phys. Rev.D88,069902(2013)].

\bibitem{Banados:1992wn}
M.~Banados, C.~Teitelboim, and J.~Zanelli, ``{The Black hole in
  three-dimensional space-time},''
  \href{http://dx.doi.org/10.1103/PhysRevLett.69.1849}{{\em Phys. Rev. Lett.}
  {\bfseries 69} (1992) 1849--1851},
\href{http://arxiv.org/abs/hep-th/9204099}{{\ttfamily arXiv:hep-th/9204099
  [hep-th]}}.

\bibitem{Awad:2000ac}
A.~M. Awad and C.~V. Johnson, ``{Scale versus conformal invariance in the AdS /
  CFT correspondence},''
  \href{http://dx.doi.org/10.1103/PhysRevD.62.125010}{{\em Phys. Rev.}
  {\bfseries D62} (2000) 125010},
\href{http://arxiv.org/abs/hep-th/0006037}{{\ttfamily arXiv:hep-th/0006037
  [hep-th]}}.

\bibitem{Jackiw:2011vz}
R.~Jackiw and S.~Y. Pi, ``{Tutorial on Scale and Conformal Symmetries in
  Diverse Dimensions},''
  \href{http://dx.doi.org/10.1088/1751-8113/44/22/223001}{{\em J. Phys.}
  {\bfseries A44} (2011) 223001},
\href{http://arxiv.org/abs/1101.4886}{{\ttfamily arXiv:1101.4886 [math-ph]}}.

\bibitem{Oz:2018yaz}
Y.~Oz, ``{On Scale Versus Conformal Symmetry in Turbulence},''
\href{http://arxiv.org/abs/1801.04388}{{\ttfamily arXiv:1801.04388 [hep-th]}}.

\bibitem{Ooguri:2016pdq}
H.~Ooguri and C.~Vafa, ``{Non-supersymmetric AdS and the Swampland},''
\href{http://arxiv.org/abs/1610.01533}{{\ttfamily arXiv:1610.01533 [hep-th]}}.

\bibitem{Hayden:2011ag}
P.~Hayden, M.~Headrick, and A.~Maloney, ``{Holographic Mutual Information is
  Monogamous},'' \href{http://dx.doi.org/10.1103/PhysRevD.87.046003}{{\em Phys.
  Rev.} {\bfseries D87} no.~4, (2013) 046003},
\href{http://arxiv.org/abs/1107.2940}{{\ttfamily arXiv:1107.2940 [hep-th]}}.

\bibitem{Vidal:2007hda}
G.~Vidal, ``{Entanglement Renormalization},''
  \href{http://dx.doi.org/10.1103/PhysRevLett.99.220405}{{\em Phys. Rev. Lett.}
  {\bfseries 99} no.~22, (2007) 220405},
\href{http://arxiv.org/abs/cond-mat/0512165}{{\ttfamily arXiv:cond-mat/0512165
  [cond-mat]}}.

\bibitem{Swingle:2009bg}
B.~Swingle, ``{Entanglement Renormalization and Holography},''
  \href{http://dx.doi.org/10.1103/PhysRevD.86.065007}{{\em Phys. Rev.}
  {\bfseries D86} (2012) 065007},
\href{http://arxiv.org/abs/0905.1317}{{\ttfamily arXiv:0905.1317
  [cond-mat.str-el]}}.

\bibitem{Czech:2015kbp}
B.~Czech, L.~Lamprou, S.~McCandlish, and J.~Sully, ``{Tensor Networks from
  Kinematic Space},'' \href{http://dx.doi.org/10.1007/JHEP07(2016)100}{{\em
  JHEP} {\bfseries 07} (2016) 100},
\href{http://arxiv.org/abs/1512.01548}{{\ttfamily arXiv:1512.01548 [hep-th]}}.

\bibitem{PhysRevLett.45.855}
Y.~Gefen, B.~B. Mandelbrot, and A.~Aharony, ``Critical phenomena on fractal
  lattices,'' \href{http://dx.doi.org/10.1103/PhysRevLett.45.855}{{\em Phys.
  Rev. Lett.} {\bfseries 45} (Sep, 1980) 855--858}.

\bibitem{Ren:2016xhb}
J.~Ren, ``{Asymptotically AdS spacetimes with a timelike Kasner singularity},''
  \href{http://dx.doi.org/10.1007/JHEP07(2016)112}{{\em JHEP} {\bfseries 07}
  (2016) 112},
\href{http://arxiv.org/abs/1603.08004}{{\ttfamily arXiv:1603.08004 [hep-th]}}.

\bibitem{Lin:2004nb}
H.~Lin, O.~Lunin, and J.~M. Maldacena, ``{Bubbling AdS space and 1/2 BPS
  geometries},'' \href{http://dx.doi.org/10.1088/1126-6708/2004/10/025}{{\em
  JHEP} {\bfseries 10} (2004) 025},
\href{http://arxiv.org/abs/hep-th/0409174}{{\ttfamily arXiv:hep-th/0409174
  [hep-th]}}.

\end{thebibliography}

\providecommand{\href}[2]{#2}\begingroup\raggedright\endgroup

\end{document}